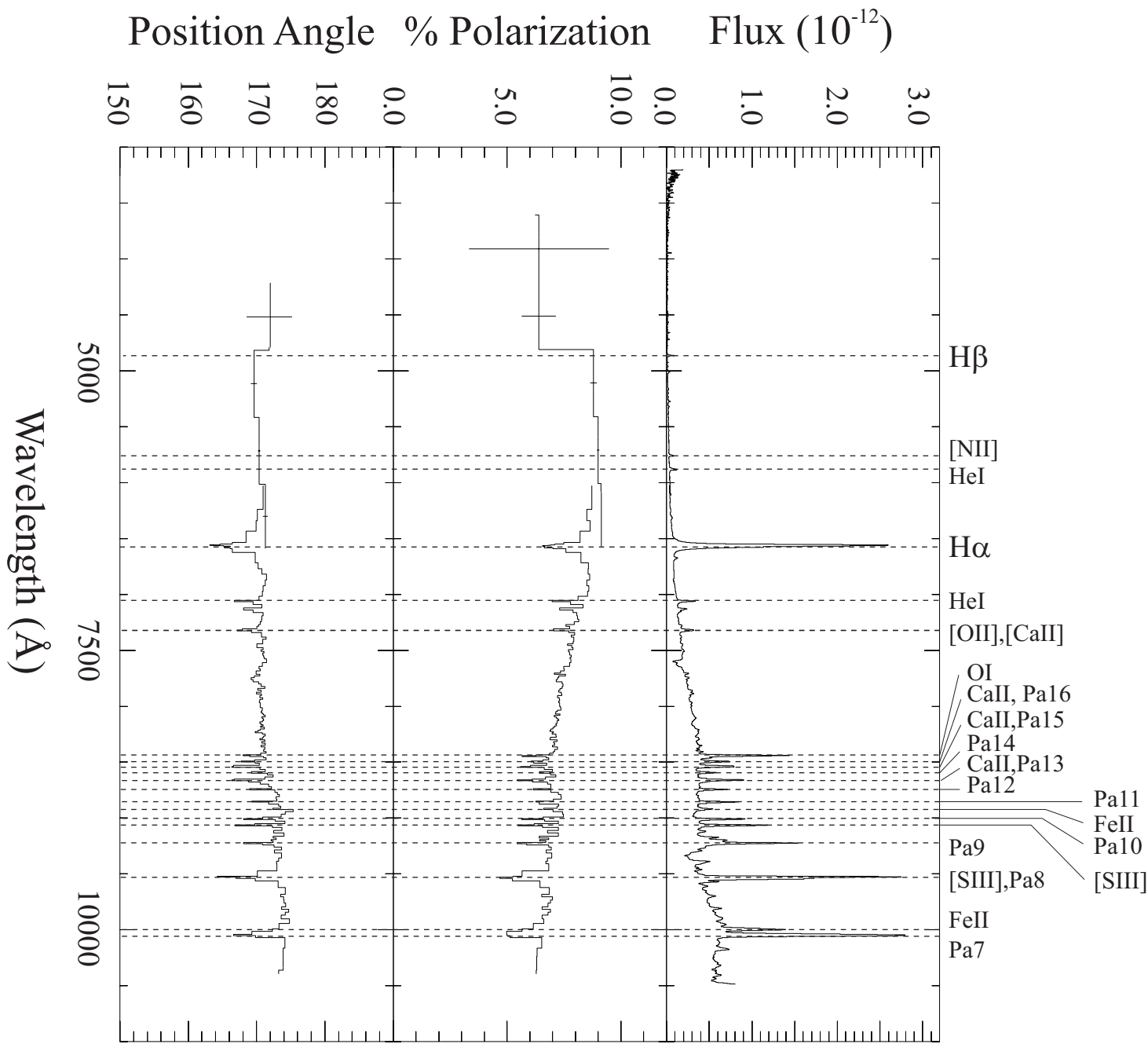

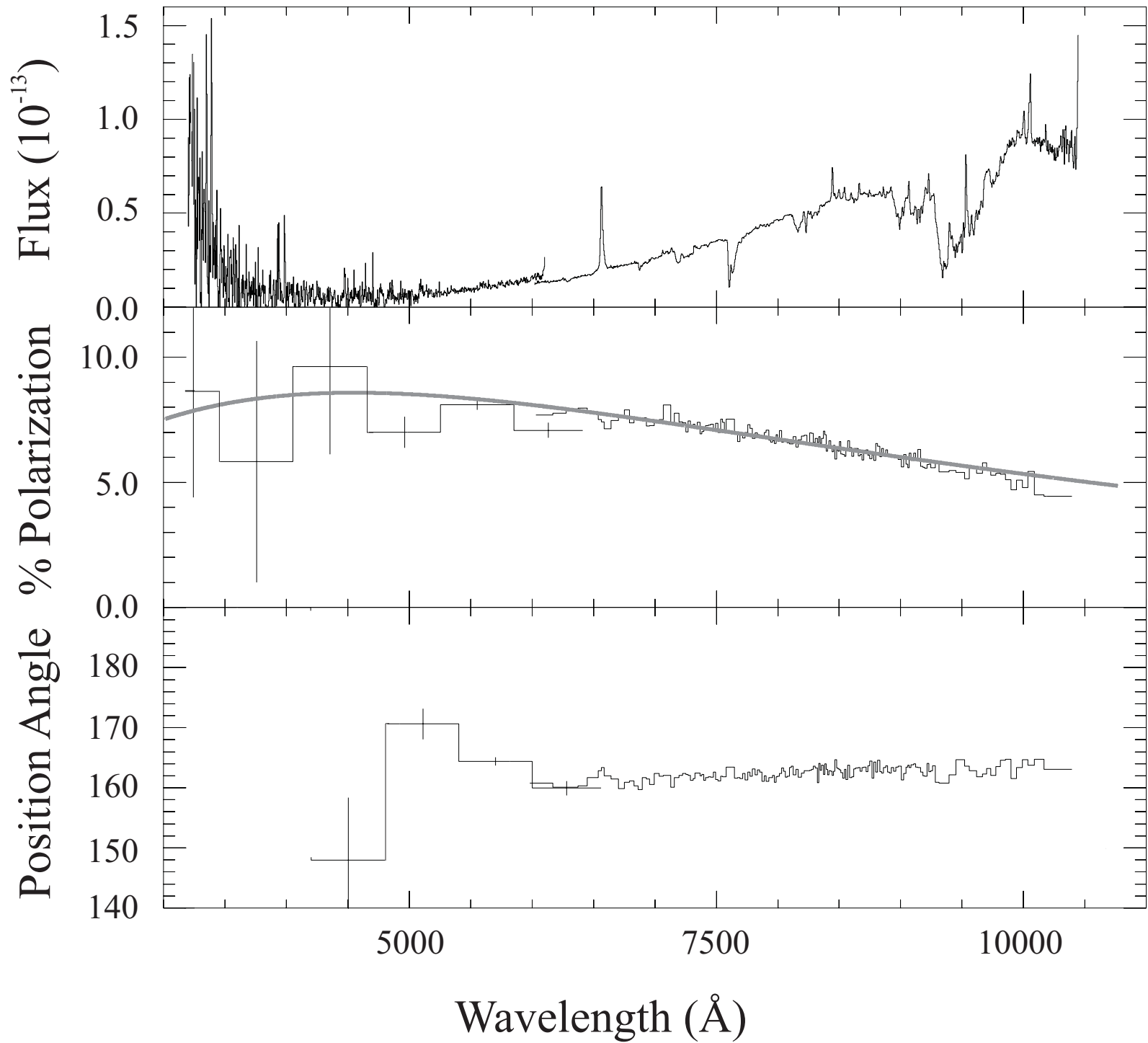

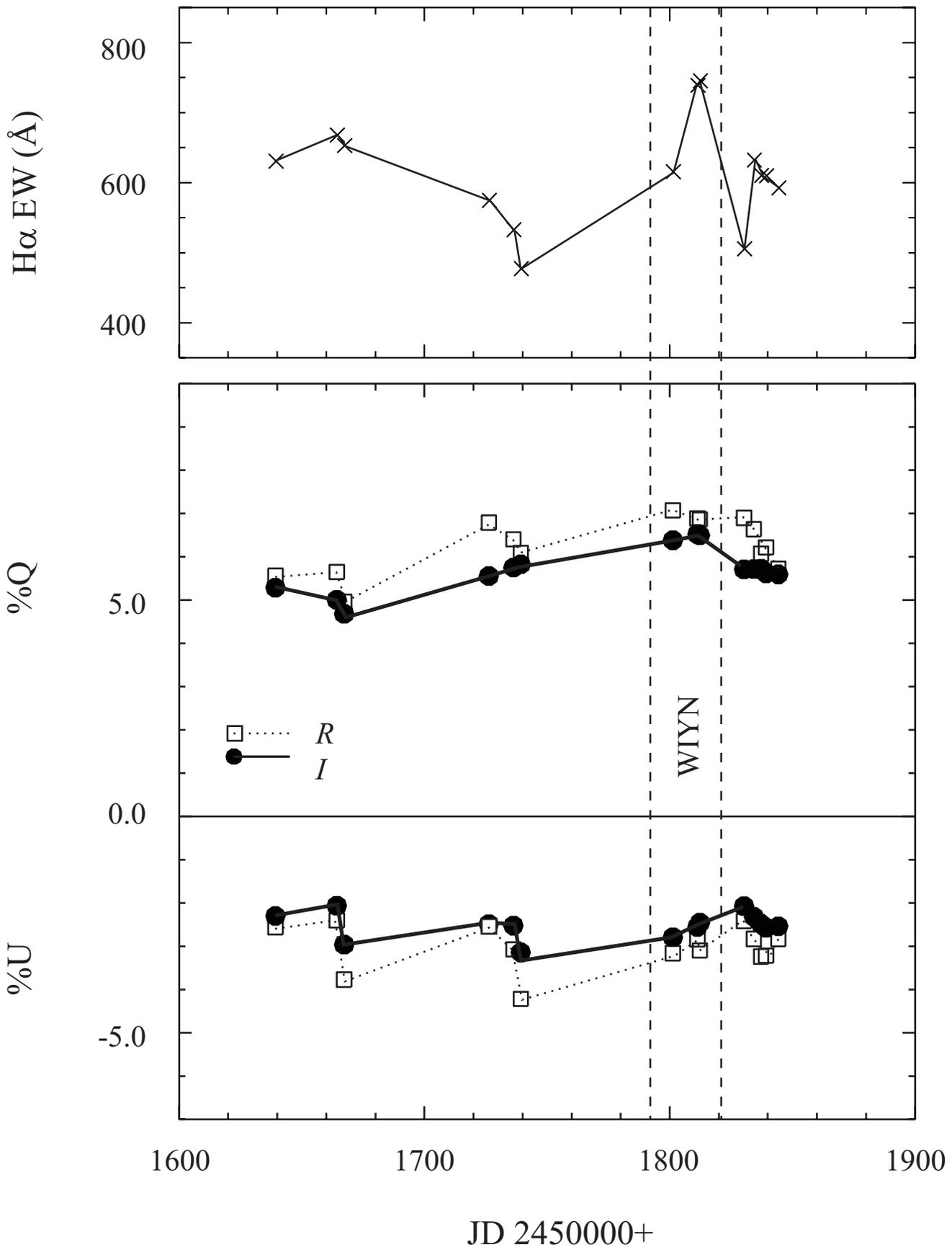

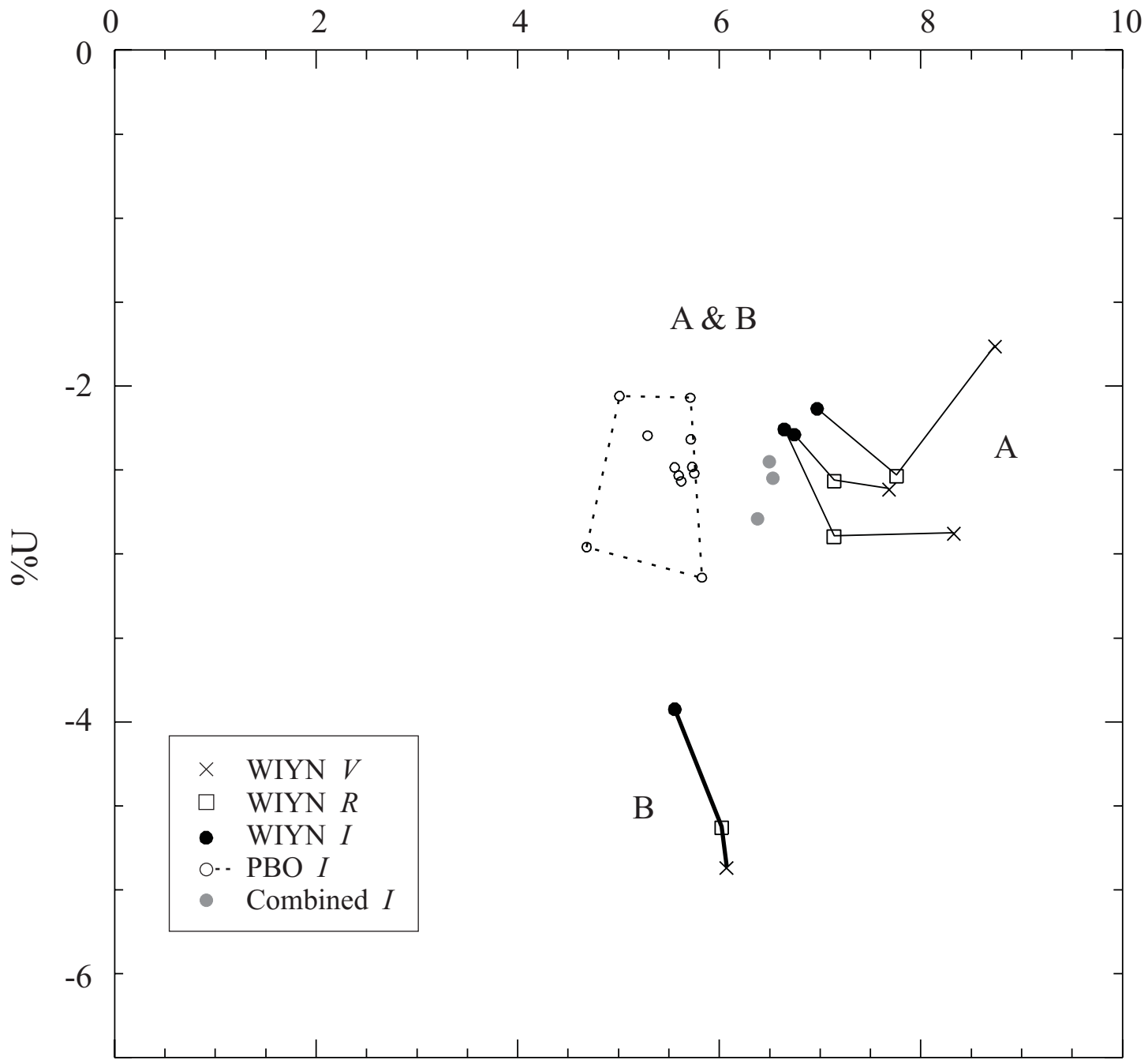

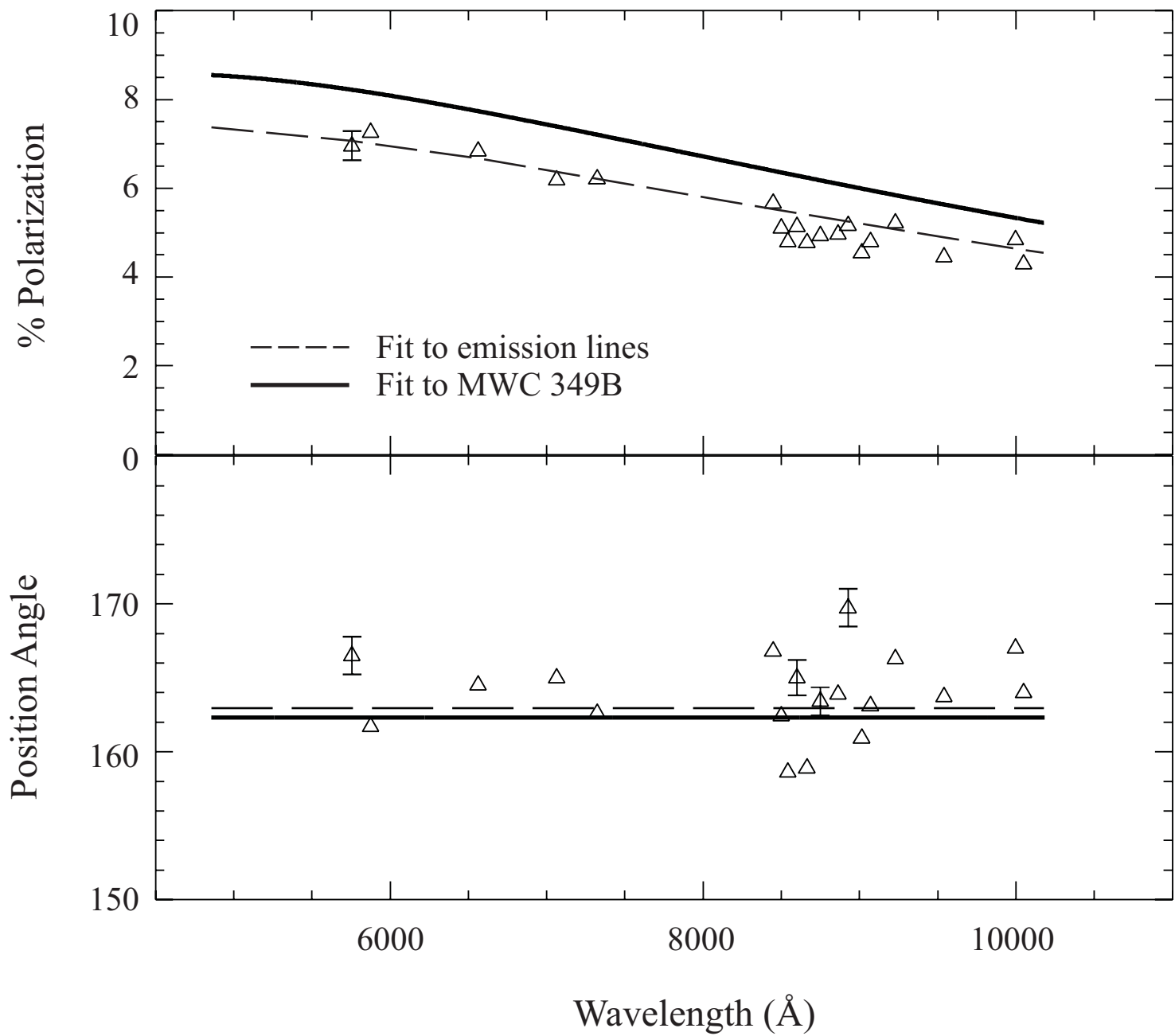

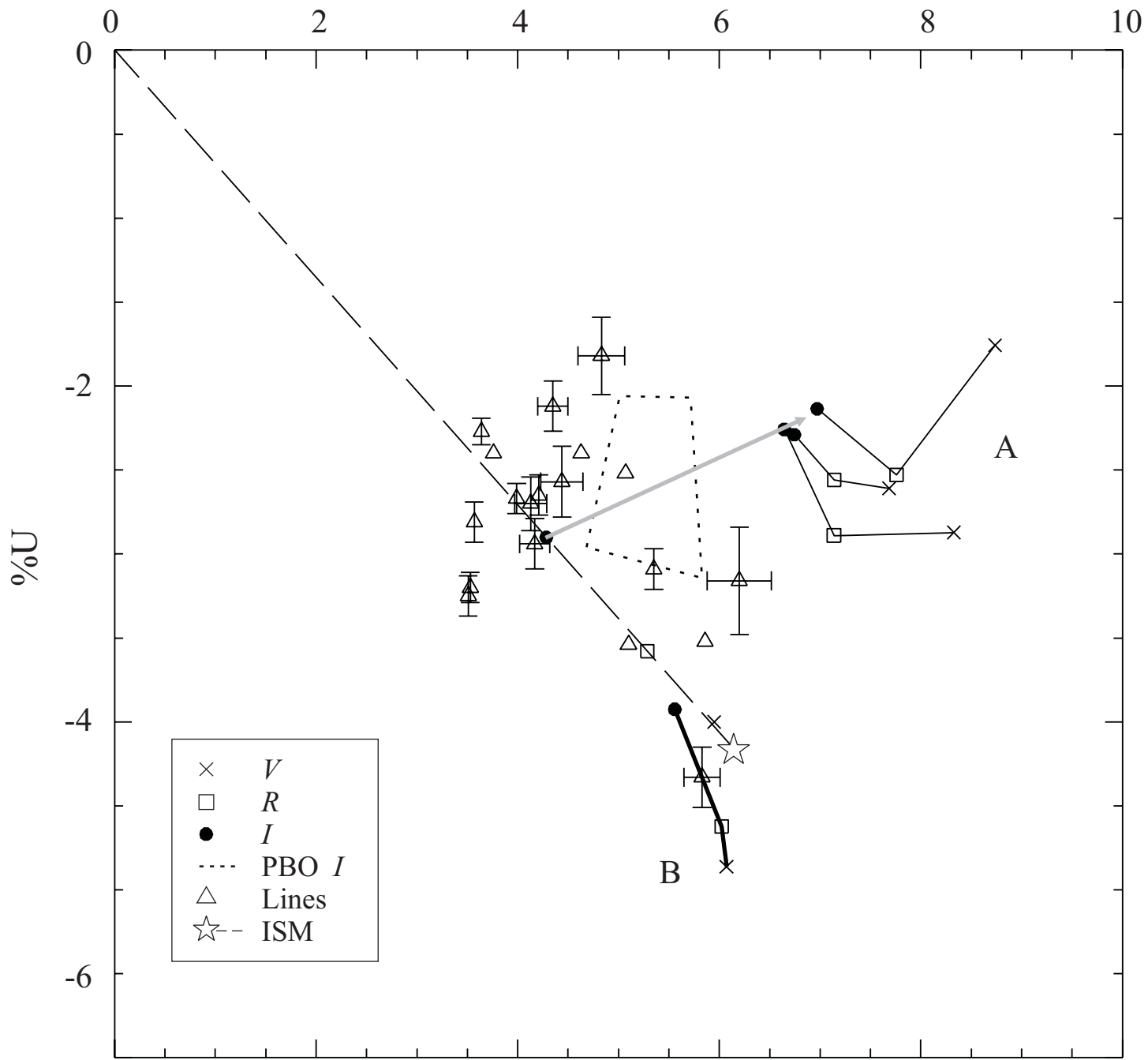

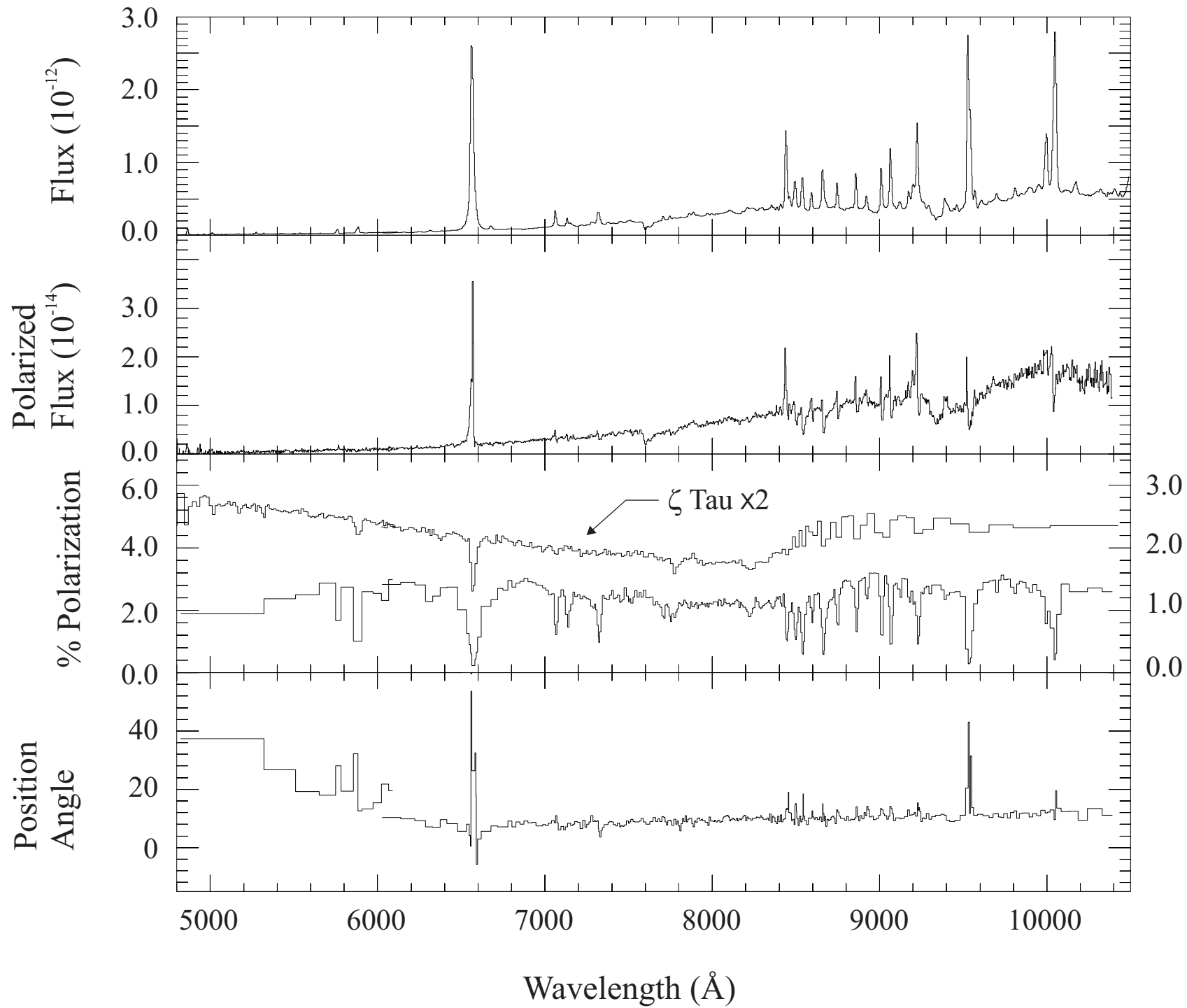

Spectropolarimetric Clues to the Structure
and Evolutionary Status of MWC 349A

Jill M. Meyer,[1,2] Kenneth H. Nordsieck[1], Jennifer L. Hoffman[1]

ABSTRACT

We present visible-wavelength spectropolarimetric measurements of the emission-line star MWC 349A and its close optical companion MWC 349B, conducted with the HPOL spectropolarimeter on both the 0.9 m telescope at the University of Wisconsin's Pine Bluff Observatory and the 3.5 m WIYN telescope on Kitt Peak. Our measurements allow us to estimate the interstellar polarization contribution and thus constrain the intrinsic polarization of MWC 349A, which we find to be consistent in position angle with the dusty disk seen perpendicular to the bipolar outflow. Our analysis reopens the possibility that MWC 349A may be part of the Cyg OB2 association, suggests that it is not a physical companion to MWC 349B, and supports the classification of MWC 349A as a pre-main sequence B[e] star.

1. INTRODUCTION

MWC 349A (V1478 Cyg, J2000.0 RA $20^h32^m45^s.53$, Dec $+40°39'37''.0$) is a variable B[e] star at an estimated distance of 1.2 kpc (Cohen et al. 1985) in the direction of the Cyg OB2 association. It is well known for emission lines as well as a strong radio continuum. Hydrogen recombination masing lines were found in 1989 (Martin-Pintado et al. 1989) and lasing lines in 1996 (Strelnitski et al. 1996); these lines were shown to originate from a circumstellar Keplerian disk that is nearly edge-on (Gordon 1992; Thum, Martin-Pintado, & Bachiller 1992). Radio observations by White & Becker (1985) revealed a bipolar outflow at a position angle of 10°, suggesting a disk position angle of 100°; another estimate by Planesas, Martin-Pintado, & Serabyn (1992) put the disk position angle at $(107 \pm 7)°$. Recently, Danchi, Tuthill, & Monnier (2001) successfully imaged a dusty disk in the system, oriented at 100° as predicted. A B0III star 2.4″ away, MWC 349B, may be a companion to the primary star (Cohen et al. 1985).

The evolutionary status of MWC 349A is unknown; it possesses characteristics of both pre-main sequence B[e] stars and evolved B[e] supergiants (Lamers et al. 1998). Earlier authors suggested other, widely varying interpretations (see the review in White & Becker 1985).

The optical broadband polarization of MWC 349A was first measured by Elvius (1974), who found it to be about 6% and wavelength-dependent. Zickgraf & Schulte-Ladbeck (1989) and Yudin (1996) confirmed Elvius's measurement. These authors attributed the high level of polarization to electron and/or dust scattering in the circumstellar disk, in addition to substantial interstellar polarization. Our new spectropolarimetric observations allow us to estimate the contribution of this interstellar polarization component; we also analyze the photometric and polarimetric variations of MWC 349A and suggest an explanatory model.

[1] Department of Astronomy, University of Wisconsin-Madison, 475 N. Charter St., Madison, WI 53706
[2] Maria Mitchell Observatory, 3 Vestal St., Nantucket, MA 02554

2. OBSERVATIONS

Using the HPOL spectropolarimeter (Nordsieck & Harris 1996), we observed MWC 349 over a period of several months in 2000 with the 0.9 m telescope at Pine Bluff Observatory (PBO) near Madison, WI. The observed wavelength range was 6000–10500 Å with a resolution of 10 Å; the slit was 6″ wide and oriented at a PA of 90°. We applied synthetic filters to the spectropolarimetric observations to obtain the polarization in the Johnson $V$, $R$, and $I$ passbands; Table 1$a$ presents these results. Since the 0.9 m telescope was not able to resolve the two stars, these measurements refer to MWC 349A and B combined.

Using the same spectropolarimeter, we also observed MWC 349A and MWC 349B separately with the 3.5 m Wisconsin-Indiana-Yale-NOAO (WIYN) telescope on Kitt Peak during September 2000[3]. The wavelength range of these observations was 4500–10500 Å, with a resolution of 7.5 Å at wavelengths below 6000 Å and 10 Å at longer wavelengths (Nordsieck & Harris 1996). The slit was 1.5″ wide and oriented at a PA of 190°, roughly perpendicular to the stellar separation. In Table 1$b$, we present the WIYN synthetic filter polarization results for each star.

All observations were done in "long-slit" mode, in which the background is measured in the same slit as, but far away from, the target. Reduction procedures were carried out using the REDUCE software package and included wavelength and flux calibration (Wolff, Nordsieck, & Nook 1996, and references therein). Several flux standard stars were observed each night, and a mean extinction curve was used to correct for atmospheric effects. The polarization was corrected for instrumental effects using a provisional calibration that is accurate to better than 0.1%. We note that since the synthetic filter routine averages over wavelength and not intensity, the emission lines do not contribute appreciably to the filter polarization results.

3. RESULTS

3.1. *Continuum Polarization*

PBO combined observations of MWC 349A and MWC 349B over nearly seven months show $R$-band polarizations of around 7% with position angles around 167°. As shown in Table 1$a$, the values of both the polarization and position angle showed temporal variability over several weeks on a scale larger than our observational errors, but with no apparent periodicity (see also Figure 3). The range in each quantity is consistent with previous measurements (Elvius 1974; Zickgraf & Schulte-Ladbeck 1989; Yudin 1996). Table 1*a* also shows that the polarization rises toward bluer wavelengths.

---

[3] The WIYN Observatory is a joint facility of the University of Wisconsin-Madison, Indiana University, Yale University, and the National Optical Astronomy Observatories.

At WIYN, we benefited from sub-arcsecond seeing, and were able to separate the two stars; these results are presented in Table 1*b*. The *R*-band polarization of MWC 349A was around 8% with a position angle of 170°; over the course of 10 days at WIYN, these values again displayed variations larger than our observational errors. Our single observation of MWC 349B yielded an *R*-band polarization of 7.5% at 161.8°. In the *R* and *I* bands, the vector differences between the polarization vector of B and the polarization vectors of each observation of A are larger than the observational errors, indicating that the two stars are distinct in their polarization. Figures 1 and 2 compare the flux, polarization, and position angle spectra of A and B. MWC 349B does not show the emission line polarization features of MWC 349A, and its polarization spectrum resembles an interstellar polarization curve; the combination of these facts suggests that its polarization is entirely interstellar in origin. (Some of the emission lines in the flux spectrum of MWC 349B component may be a result of observational contamination from the A component.) A modified Serkowski law fit (Serkowski, Mathewson, & Ford 1975; Wilking, Lebofsky, & Rieke 1982) to the polarization spectrum of MWC 349B yielded an interstellar polarization estimate of $P_{max} = (8.58 \pm 0.05)\%$, $\theta = (162.4 \pm 0.2)°$, and $\lambda_{max} = (4546 \pm 28)$ Å (see §4 for discussion).

We artificially combined each of the WIYN observations of MWC 349A with the single WIYN observation of MWC 349B by dividing the sum of the two polarized flux spectra by the sum of the two flux spectra; the resulting polarization spectrum in each case was consistent with the average polarization spectrum observed at PBO. Figure 3 displays the polarization variations in the PBO and artificially-combined WIYN observations in the *R* and *I* bands. (We do not include the *V*-band results here because we consider the PBO data to be somewhat unreliable in this band due to the faintness of the star.) The observed variations could be due to the variation in polarization of the A component, the variation in relative flux between the A and B components, or some combination of these effects. Relative flux variations should result in changes in the equivalent width of the Hα line, which we indeed observe (Figure 3, top panel). However, the fact that the changes in Hα equivalent width do not correlate with the changes in polarization argue that some intrisic polarization variation must also occur.

Our separate WIYN observations of the two components corroborate this conclusion. Figure 4 compares the resolved WIYN *VRI* continuum observations of MWC 349A and MWC 349B with the PBO *I*-band observations of the combined system on a *Q-U* diagram. The artificially combined WIYN *I*-band results are also shown. It can be demonstrated that if the variations in the combined observations were due only to changes in the relative flux of the two components, all PBO points would lie along a line between the *I*-band point of MWC 349B and the *I*-band points of the three MWC 349A observations. The three WIYN combined points lie along this line by construction. However, the fact that the PBO observations deviate significantly from the A-B line indicates that the intrinsic polarization of MWC 349A is indeed variable on a timescale of months, and that this variability is the dominant contributor to the variability of the combined polarization. From Figure 4, we see that these intrinsic variations have an amplitude of a few percent. To quantify this amplitude more precisely, we need an estimate of the interstellar polarization contribution. We will make such an estimate in §4 below, and revisit these continuum polarization results in §5.

3.2 *Line Polarization*

Figure 1 shows that the emission lines in the spectrum of MWC 349A are polarized differently from the continuum, another indicator that the A component possesses intrinsic polarization. For each of our three WIYN observations, we calculated the polarization values of the strongest emission lines, using the flux-equivalent-width method described in Hoffman, Nordsieck, & Fox (1998). In Table 2, we identify the lines and report the results. We used the error-weighted means of the *Q* and *U* Stokes parameters to calculate weighted average *P* and PA values; equivalent widths were simply averaged. Lines toward the blue end of the spectrum are more highly polarized than red lines, while the position angle remains roughly constant near 165° for all lines.

To investigate the corresponding variability of the line strengths, we chose strong, isolated emission lines of three different species to trace through our three WIYN observations of MWC 349A: OI $\lambda$8447, [SIII] $\lambda$9069, and Pa7 $\lambda$10049. Table 3 tabulates the results, which show that the equivalent widths of these three emission lines remain nearly constant with time, while the continuum flux measured at 9754 Å, a nearby region uncontaminated by emission lines, varies significantly. (Though we have no independent method of estimating errors in our equivalent width measurements, our experience with the data from this instrument indicates that these values are consistent with a nonvarying equivalent width for each line.) For this to occur, the line strengths must be varying in step with the continuum, suggesting that both the line- and continuum-forming regions are occasionally blocked from view by occulting material. However, our observations show that the polarization does not vary as quickly as the continuum flux; it changes instead on a much longer timescale (Figure 3). The occulting material must therefore block light without changing the overall polarization.

Yudin (1996) also concluded that the polarization and brightness of MWC 349A do not vary together; however, he found the intrinsic polarization to vary widely on a timescale of hours, and the brightness to have slow variations. We have not observed these quick polarization variations (Table 1*b*); see §5 for a discussion of this discrepancy.

4. INTERSTELLAR POLARIZATION

We expect that the polarization of the interstellar medium in the direction of MWC 349A contributes significantly to the polarization we observe. Not only does the observed continuum polarization have a large magnitude, but its position angle is not consistent with the observed position of either the disk or the outflow. Both these circumstances suggest that interstellar dust is a substantial polarizing mechanism along the line of sight to MWC 349.

Yudin (1996) estimated the interstellar polarization toward MWC 349A by combining polarization measurements made by Whittet et al. (1992) of 12 stars in the Cyg OB2 association;

he found $P_{max}$ = (3.5 ± 0.5)% and θ = (100 ± 20)°. Our spectropolarimetric observations of the system allow us to make an independent estimate. As described in §3.1, we first attempted to do so by fitting a modified Serkowski law to the polarization spectrum of MWC 349B. However, when we subtracted this Serkowski fit ($P_{max}$ = 8.58%, θ = 162.4°, $\lambda_{max}$ = 4546 Å) from the polarization spectrum of MWC 349A, we found that all the emission lines were left with unusually high intrinsic polarization values (~ 1%). We looked for differences in this intrinsic polarization between lines of different atomic species, but could find no pattern. Figure 5 shows the observed line polarization values and the interstellar fit to MWC 349B (*solid line*). The differences between the individual points and the solid line illustrate the large and seemingly random intrinsic line polarizations implied by this interstellar estimate. We believe this result to be unrealistic, and have therefore sought another method of estimating the interstellar polarization.

In Be stars, emission lines have been observed to be less polarized than the continuum (Coyne 1976). Line emission produced in the disk is scattered less (and therefore polarized less) because it travels a shorter distance and through a less dense area than the continuum radiation. As a result, emission lines may have little to no intrinsic polarization, and therefore any significant polarization exhibited by emission lines is usually attributed to interstellar effects. Making the same assumption in the case of MWC 349A, we have fit a modified Serkowski law to the values of the emission line polarizations listed in Table 2, finding $P_{max}$ = (7.38 ± 0.11)%, θ = (164.2 ± 0.4)°, $\lambda_{max}$ = 4500 Å. (Based on the line polarization results, $\lambda_{max}$ must be smaller than 5754 Å; we did not fit this parameter, but rather chose a reasonable value near that of the MWC 349B interstellar estimate.) The dashed lines in Figure 5 show the resulting fit. Removing this new interstellar estimate from our WIYN observations of MWC 349A yielded the intrinsic polarization values displayed in Table 4. (We have omitted the *B*-band results, as these all had very large errors.) The derived intrinsic position angles fall approximately 90° from the orientation of the observed disk (100°; Danchi et al. 2001), suggesting that the intrinsic polarization of MWC 349A does indeed arise in this disk. We note that while Yudin's (1996) interstellar estimate required him to propose two circumstellar disks at different orientation angles in MWC 349A, our new estimate is consistent with the simpler scenario in which polarization arises in the single observed disk. This agreement with previous findings reinforces our confidence in this estimate.

Figure 6 updates Figure 4 with the polarization values of the emission lines and the interstellar polarization fit to these values. The direction of MWC 349A's intrinsic polarization is illustrated by the gray arrow connecting the *I* marker along the interstellar line with the *I*-band observational points. The scatter in the line polarization values around the interstellar fit is mostly in the direction of this intrinsic polarization, possibly indicating the presence of intrinsic polarization in the emission lines. (For example, OI λ8447 may be intrinsically polarized if it is pumped by Lyβ photons.) This is, of course, contrary to our initial assumption, and if the lines are indeed intrinsically polarized, our estimated interstellar position angle may be too large, i.e., the interstellar line in Figure 6 should in fact be rotated more toward the observation of MWC 349B. However, a comparison of the MWC 349B observation with the *VRI* points along the interstellar line makes it clear that even in this case, the polarization of MWC 349B would be significantly different from that of the interstellar estimate.

5. DISCUSSION AND CONCLUSIONS

The polarization of the PBO combined observations (enclosed by the dotted bubble in Figure 6) also varies approximately in the direction of MWC 349A's intrinsic polarization, confirming our conclusion (in §3.1) that the polarization variation in A causes most of the variation in the combined observations. Since we now know the magnitude and direction of the interstellar component, we know that at the time of the PBO observations, the intrinsic polarization of MWC 349A was smaller in magnitude than it was at the time of the WIYN observations. We can also see that MWC 349A varied in intrinsic polarization by a factor of 2 to 3 over the entire observational period.

Removing our interstellar estimate from our WIYN observations of MWC 349A leaves the intrinsic spectra shown in Figure 7, where we have combined the three WIYN observations via error-weighted vector addition. If our assumption that the emission lines are intrinsically unpolarized is correct, the lines should not appear in the polarized flux spectrum. Figure 7 shows that some are still visible, though fairly weak. This is partly due to problems with data extraction near strong lines (especially apparent near H$\alpha$ and [SIII]/Pa8), but may also indicate small residual line polarization effects, as discussed in §4 above. The intrinsic polarization spectrum, shown in the third panel, strongly resembles that of the Be star $\zeta$ Tau plotted above it. The magnitude of the polarization is large, around 2%, and the Paschen jump is clearly defined. These similarities to Be star characteristics suggest that although the outer circumstellar environment of MWC 349A is quite dusty (Cohen et al. 1985; Danchi et al. 2001), electron scattering is the primary polarizing mechanism in MWC 349A.

It is notable that MWC 349B is more highly polarized than most of the emission lines of MWC 349A, and also more highly polarized than our interstellar polarization estimate (Figure 6). Either MWC 349B has its own intrinsic polarization, or it is not associated with A. Subtracting our interstellar estimate from our observation of MWC 349B, we find that its residual polarization is around 1.6%, quite high for an intrinsic polarization level, especially considering this star's lack of unique characteristics and spectral features. Since the polarization spectrum of MWC 349B is so well fit by a Serkowski law (Figure 2), we believe B's polarization is all interstellar. This implies that is is not in fact a companion to MWC 349A, but lies farther from us, with a correspondingly higher contribution from interstellar polarization. Statistically, this is a feasible scenario: from the photometric study of the Cyg OB2 region by Massey & Thompson (1991), we estimate that the mean separation between stars in the neighborhood of MWC 349A brighter than $m_V = 15$ is ~ 2.6′. Given this high stellar surface density, the probability of a chance alignment between A and B at the observed 2.4″ separation is around 1 in 1300, while the probability of such an alignment occurring for *any* of the stars in and around Cyg OB2 is close to 1 in 3.

Kobulnicky, Molnar, & Jones (1994) used polarization observations of 132 members of the Cyg OB2 association to create a polarization map of the region. Several stars in their sample near MWC 349 have polarization and position angle characteristics similar to those of our interstellar estimate. Based on this similarity, we suggest that in fact MWC 349A is part of Cyg OB2, a possibility previously proposed by Braes, Habing, & Schoenmaker (1972); this would place it at a distance of ~ 1.7 kpc (Massey & Thompson 1991). This conclusion is supported by a recent study by Knödlseder (2000), which found that Cyg OB2 is much larger than previously thought, and that MWC 349A lies within the boundaries of the association.

Cohen et al. (1985) disputed the idea that MWC 349A was a member of Cyg OB2; they cited radio emission apparently connecting A and B and used their classification of B as a B0III star, combined with the high extinction apparent in the spectrum of A, to derive a distance of ~ 1.2 kpc. However, the radio arc observed by Cohen et al. does not unambiguously link the two stars. If associated with A, the arc could instead be caused by the collision of A's wind with that of another star, not B, or with a region of dense interstellar material. But the arc may not even be associated with A; though its position and morphology suggest an association, it is clearly separated from the rest of A's emission contours. Larger radio maps of the surrounding region by Wendker (1984) and Wendker, Higgs, & Landecker (1991) revealed complex structure on a variety of scales; the arc in question could be part of this background structure. In short, we believe the radio data leave open the possibility that the two stars are not physical companions. If this is the case, the dusty circumstellar environment of MWC 349A must contribute heavily to its extinction. Because the position angle of MWC 349B's polarization is so close to that of our interstellar estimate for MWC 349A (Figure 5), however, we consider it likely that MWC 349B is also a member of Cyg OB2; it could lie behind the dusty region around MWC 349A, or perhaps on the farther edge of the association.

If the apparently evolved MWC 349B is not a companion to MWC 349A, then it becomes more likely that A is a pre-main sequence B[e] star rather than an evolved B[e] supergiant. MWC 349A's rotating disk (Rodriguez & Bastian 1994; Danchi et al. 2001) and spectroscopic similarities to young stellar objects (Hamann & Simon 1986, 1988) also argue for a pre-main sequence classification. In addition, MWC 349A possesses several of the criteria required for membership in the HAeB[e] subclass: B[e] spectral characteristics, an association with nebulosity, rapid photometric variations, and dust signatures (Lamers et al. 1998). Its luminosity ($3 \times 10^4$ $L_o$ as derived by Cohen et al. 1985) places it on the border between the HAeB[e] ($\log L/L_o \lesssim 4.5$) and sgB[e] ($\log L/L_o \gtrsim 4.0$) subclasses (Lamers et al. 1998), and our increased distance estimate could of course increase this value. However, Cohen et al. solved for the luminosity based on the assumption that MWC 349A's intrinsic colors are like those of a normal hot star. Different spectral characteristics produce different extinction values and lead to different luminosity estimates. In particular, an $A_V$ of ~ 8.8, which Cohen et al. derived for a spectrum of the form $F_\nu \propto \nu^{1/3}$, would essentially cancel out the increase in luminosity introduced by our increased distance estimate. Overall, then, we believe our results strengthen the case for classifying MWC 349A as a pre-main sequence B[e] star.

Our observations reveal that the polarization of MWC 349A is disk-like and consistent with scattering by electrons. The fluxes of the emission lines and the continuum vary simultaneously and rapidly, while the polarization varies much more slowly. We speculate that moving clumps within the electron-scattering disk could account for the slow changes in polarization. Because the rapid photometric variations occur simultaneously in both the line- and continuum-forming regions and do not correlate with rapid changes in polarization, they presumably occur on a smaller physical scale inside this disk. They could be due to brightness variations of the star itself, or to episodic interactions between the star and disk, as long as the disk illumination remains roughly symmetric. The quick polarization variations Yudin (1996) observed could have been produced by changes in the illumination of the inner disk, perhaps due to starspots or other asymmetric stellar activity. In this picture, the electron-scattering disk lies inside and at the same position angle as the dust disk seen by Danchi et al. (2001), which presumably produces the high extinction toward MWC 349A.

We found that the polarization of the emission lines of MWC 349A decreases with increasing wavelength. If our assumption that the lines are intrinsically unpolarized is correct, then any emission lines redward of the region where the interstellar polarization law drops to zero (around 3 µm) should be unpolarized. Infrared spectropolarimetry of sufficient resolution would then allow a more reliable estimate of the position angle of interstellar polarization via the difference between the polarizations of the visual and infrared emission lines. Some infrared spectropolarimetry with a wavelength range of 8-13 µm has been published by Aitken et al. (1990), but their spectral resolution was not high enough to resolve emission lines.

It would also be interesting to use high-resolution spectropolarimetry to obtain the polarization across the emission line profiles. Hamann & Simon (1988) used emission line profiles in the flux spectrum to locate emitting regions in the disk and outflow; observation of polarized line profiles could provide independent confirmation of their results. Such observations in conjunction with photometric and polarimetric monitoring could also address the question of whether the polarimetric variations are due to motions of clumps within the disk or changes in the symmetry of the stellar illumination.

We thank J. S. Gallagher, H. A. Kobulnicky, V. Strelnitski, J. Mathis, and an anonymous referee for helpful comments on this paper. J. M. Meyer thanks V. Strelnitski for his collaborative efforts and B. L. Babler for assistance with the data reduction, and gratefully acknowledges support from NSF/REU grant AST-9820555 and the Maria Mitchell Observatory.

REFERENCES

Aitken, D. K., Smith, C. H., Roche, P. F., & Wright, C. M. 1990, MNRAS, 247, 466
Braes, L. L. E., Habing, H. J., & Schoenmaker, A. A. 1972, Nature, 240, 230
Cohen, M., Bieging, J. H., Dreher, J. W., & Welch, W. J. 1985, ApJ, 292, 249
Coyne, G. V. 1976, A&A, 49, 89
Danchi, W. C., Tuthill, P. G., & Monnier, J. D. 2001, ApJ, 562, 440
Elvius, A. 1974, A&A, 34, 371
Gordon, M. A. 1992, ApJ, 387, 701
Hamann, F. & Simon, M. 1986, ApJ 311, 909
Hamann, F. & Simon, M. 1988, ApJ, 327, 876
Hoffman, J. L., Nordsieck, K. H., & Fox, G. K. 1998, AJ, 115, 1576
Knödlseder, J. 2000, A&A, 360, 539
Kobulnicky, H. A., Molnar, L. A., & Jones, T. J. 1994, AJ, 107, 1433
Lamers, H. J. G. L. M.; Zickgraf, F.-J., de Winter, D., Houziaux, L., & Zorec, J. 1998, A&A, 340, 117
Martin-Pintado, J., Bachiller, R., Thum, C., & Walmsley, M. 1989, A&A, 215, L13
Massey, P., & Thompson, A. B. 1991, AJ, 101, 1408
Nordsieck, K. H. & Harris, W. 1996, in ASP Conf. Ser. 97, Polarimetry of the Interstellar Medium, ed. W.G. Roberge & D.C.B. Whittet (San Francisco: ASP), 100
Planesas, P., Martin-Pintado, J., & Serabyn, E. 1992, ApJ, 386, L93
Rodriguez, L. F., & Bastian, T. S. 1994, ApJ, 428, 324
Serkowski, K., Mathewson, D. S., & Ford, V. L. 1975, ApJ, 196, 261
Strelnitski, V., Haas, M., Smith, H., Erickson, E., Colgan, S., & Hollenback, D. 1996, Science, 272, 1459
Thum, C., Martin-Pintado, J., & Bachiller, R. 1992, A&A, 256, 507
Wendker, H. J. 1984, A&AS, 58, 291
Wendker, H. J., Higgs, L. A., & Landecker, T. L. 1991, A&A, 241, 551
White, R., and Becker, R. 1985, Ap.J, 297, 677
Wilking, B. A., Lebofsky, M. J., & Rieke, G. H. 1982, AJ, 87, 695
Whittet, D. C. B., Martin, P. G., Hough, J. H., Rouse, M. F., Bailey, J. A., & Axon, D. J. 1992, ApJ, 386, 562
Wolff, M. J., Nordsieck, K. H., & Nook, M. A. 1996, AJ, 111, 856
Yudin, R.V. 1996, A&A, 312, 234
Zickgraf, F.-J., and Schulte-Ladbeck, R. E. 1989, A&A, 214, 274

FIG. 1.—*From top*: Flux (10⁻¹² ergs cm⁻² s⁻¹ Å⁻¹), percent polarization, and position angle (degrees) spectra for the 2000 September 25 WIYN observation of MWC 349A. The Stokes parameters producing the polarization and position angle are binned to constant internal errors of 0.17% (3000-6000 Å) and 0.09% (6000-10500 Å). The strongest emission lines have been identified.

FIG. 2.—As in Fig. 1, for the 2000 September 11 WIYN observation of MWC 349B. Stokes parameters are binned to 600Å for 3000-6000 Å and to a constant internal error of 0.2% for 6000-10500 Å. In the polarization (*middle panel*), the curved line represents a best-fit interstellar polarization curve (§3.1).

FIG. 3.—Variations with time in the equivalent width of Hα and the Stokes parameters %Q and %U of the continuum of MWC 349A and MWC 349B combined. WIYN observations of MWC 349A have been separately combined with the single WIYN observation of MWC 349B for comparison with the PBO points (§3.1). Error bars are shown in %*Q* and %*U* when larger than the symbol. Since the equivalent width variations do not correlate with the polarization variations, we conclude some intrinsic polarization variability must exist in MWC 349A.

FIG. 4.—*Q-U* diagram comparing PBO and WIYN observations of MWC 349. The three light solid lines (*labeled "A"*) represent our WIYN observations of MWC 349A, while the thick line (*labeled "B"*) represents MWC 349B. In each case, the X marks the *V*-band polarization, the open square marks the *R* band, and the black circle marks the *I* band. *I*-band observations of MWC 349A and B combined are shown as isolated circles and labeled "A & B." The dotted bubble encloses unresolved PBO *I*-band observations of A and B combined (*open circles*), while the gray circles represent the artificially combined WIYN *I*-band observations (§3.1). In all cases, error bars are comparable in size to or smaller than the symbols. The PBO observations differ substantially from the line traced by the WIYN combined observations, providing another indication that the polarization variations in Fig. 3 are due to intrinsic polarization variations in MWC 349A and not to changes in the relative fluxes of A and B.

FIG. 5—Percent polarization (*top*) and position angle (*bottom*) of the emission lines of MWC 349A. Error bars are shown when larger than the symbol. The solid line is the best-fit interstellar polarization curve to MWC 349B shown in Fig. 2; taking this for the interstellar estimate of MWC 349A leaves the emission lines with large intrinsic polarization values that are difficult to explain. The dashed line traces the interstellar polarization curve that best fits the emission line polarization, and which we have taken for our interstellar polarization estimate, assuming the lines are intrinsically unpolarized (§4).

FIG. 6—*Q-U* diagram showing our PBO and WIYN observations with our adopted interstellar polarization estimate. The solid lines, dotted bubble, and *VRI* observations of MWC 349A and B are as in Fig. 4. Triangles represent the emission lines from Fig. 5, with error bars shown when larger than the symbol. The star represents our estimate of the magnitude of the interstellar polarization, and the dashed line defines its position angle (§4). The central wavelengths of the *V*, *R*, and *I* bands have been marked along the interstellar line (*circle*, *square, and X,*

*respectively*). The intrinsic polarization of MWC 349A in the *I* band is shown with a gray arrow. The scatter in the emission line polarization is roughly in the direction of the gray arrow, indicating that some of the lines may be intrinsically polarized.

FIG. 7— *From top*: Flux ($10^{-12}$ ergs cm$^{-2}$ s$^{-1}$ Å$^{-1}$), polarized flux (percent polarization times flux, $10^{-14}$ ergs cm$^{-2}$ s$^{-1}$ Å$^{-1}$), percent polarization, and position angle (degrees) spectra for the three WIYN observations of MWC 349A, combined via error-weighted vector addition and corrected for interstellar polarization (§4). In the polarization plot, we also show a polarization spectrum of the Be star ζ Tau, observed at PBO on 1999 March 4. For clarity, the ζ Tau spectrum is multiplied by a factor of two; its scale is shown on the right-hand axis. The Stokes parameters producing the polarization and position angle of MWC 349A are binned to constant internal errors of 0.13% (4800-6000 Å) and 0.09% (6000-10500 Å). The ζ Tau spectrum is binned to a constant internal error of 0.023%. The similarity between the polarization spectra of MWC 349A and ζ Tau suggests that electron scattering is the primary polarization mechanism in MWC 349A.

TABLE 1a
PBO OBSERVATIONS OF MWC 349A AND B COMBINED

| Date | MJD | %$P$ | $\sigma_P$ (%) | PA (°) | $\sigma_{PA}$ (°) |
|---|---|---|---|---|---|
| | | $V$ | | | |
| 2000 Apr 5  | 1639 | 6.24  | 0.25 | 164.5 | 1.2 |
| 2000 Apr 30 | 1664 | 6.49  | 0.31 | 168.4 | 1.4 |
| 2000 May 3  | 1667 | 5.79  | 0.26 | 157.4 | 1.3 |
| 2000 Jul 1  | 1726 | 10.50 | 0.83 | 179.3 | 2.3 |
| 2000 Jul 11 | 1736 | 7.18  | 1.10 | 163.6 | 4.5 |
| 2000 Jul 14 | 1739 | 8.41  | 1.10 | 162.3 | 3.8 |
| 2000 Oct 13 | 1830 | 9.43  | 0.36 | 168.3 | 1.1 |
| 2000 Oct 17 | 1834 | 7.74  | 0.23 | 172.0 | 0.9 |
| 2000 Oct 20 | 1837 | 7.55  | 0.19 | 166.0 | 0.7 |
| 2000 Oct 22 | 1839 | 8.90  | 0.20 | 173.1 | 0.6 |
| 2000 Oct 27 | 1844 | 5.95  | 0.18 | 170.2 | 0.9 |
| | | $R$ | | | |
| 2000 Apr 5  | 1639 | 6.12 | 0.05 | 167.6 | 0.2 |
| 2000 Apr 30 | 1664 | 6.14 | 0.08 | 168.5 | 0.4 |
| 2000 May 3  | 1667 | 6.23 | 0.05 | 161.4 | 0.2 |
| 2000 Jul 1  | 1726 | 7.24 | 0.15 | 169.8 | 0.6 |
| 2000 Jul 11 | 1736 | 7.10 | 0.20 | 167.2 | 0.8 |
| 2000 Jul 14 | 1739 | 7.40 | 0.21 | 162.7 | 0.8 |
| 2000 Oct 13 | 1830 | 7.31 | 0.09 | 170.4 | 0.4 |
| 2000 Oct 17 | 1834 | 7.22 | 0.04 | 168.0 | 0.2 |
| 2000 Oct 20 | 1837 | 6.88 | 0.04 | 166.0 | 0.2 |
| 2000 Oct 22 | 1839 | 7.00 | 0.04 | 166.3 | 0.2 |
| 2000 Oct 27 | 1844 | 6.39 | 0.04 | 166.9 | 0.2 |
| | | $I$ | | | |
| 2000 Apr 5  | 1639 | 5.77 | 0.02 | 168.3 | 0.1 |
| 2000 Apr 30 | 1664 | 5.42 | 0.03 | 168.8 | 0.2 |
| 2000 May 3  | 1667 | 5.54 | 0.02 | 163.9 | 0.1 |
| 2000 Jul 1  | 1726 | 6.09 | 0.03 | 168.0 | 0.1 |
| 2000 Jul 11 | 1736 | 6.28 | 0.03 | 168.2 | 0.2 |
| 2000 Jul 14 | 1739 | 6.62 | 0.03 | 165.8 | 0.1 |
| 2000 Oct 13 | 1830 | 6.08 | 0.03 | 170.0 | 0.1 |
| 2000 Oct 17 | 1834 | 6.17 | 0.02 | 169.0 | 0.1 |
| 2000 Oct 20 | 1837 | 6.25 | 0.01 | 168.3 | 0.1 |
| 2000 Oct 22 | 1839 | 6.18 | 0.02 | 167.7 | 0.1 |
| 2000 Oct 27 | 1844 | 6.14 | 0.02 | 167.8 | 0.1 |

TABLE 1*b*
WIYN OBSERVATIONS OF MWC 349A AND MWC 349B

| Target | Date | MJD | %$P$ | $\sigma_P$ (%) | PA (°) | $\sigma_{PA}$ (°) |
|---|---|---|---|---|---|---|
| | | *B* | | | | |
| MWC 349A | 2000 Sep 24 | 1811 | 6.349 | 0.232 | 169.96 | 1.05 |
| | | *V* | | | | |
| MWC 349A | 2000 Sep 14 | 1801 | 8.910 | 0.042 | 174.32 | 0.14 |
| | 2000 Sep 24 | 1811 | 8.117 | 0.045 | 170.61 | 0.16 |
| | 2000 Sep 25 | 1812 | 8.808 | 0.030 | 170.49 | 0.10 |
| MWC 349B | 2000 Sep 11 | 1798 | 9.768 | 1.803 | 174.03 | 5.29 |
| | | *R* | | | | |
| MWC 349A | 2000 Sep 14 | 1801 | 8.162 | 0.076 | 170.98 | 0.27 |
| | 2000 Sep 24 | 1811 | 7.586 | 0.070 | 170.14 | 0.26 |
| | 2000 Sep 25 | 1812 | 7.702 | 0.006 | 168.98 | 0.02 |
| MWC 349B | 2000 Sep 11 | 1798 | 7.513 | 0.022 | 161.82 | 0.08 |
| | | *I* | | | | |
| MWC 349A | 2000 Sep 14 | 1801 | 7.291 | 0.008 | 171.48 | 0.03 |
| | 2000 Sep 24 | 1811 | 7.123 | 0.004 | 170.62 | 0.02 |
| | 2000 Sep 25 | 1812 | 7.019 | 0.004 | 170.62 | 0.02 |
| MWC 349B | 2000 Sep 11 | 1798 | 6.669 | 0.011 | 162.50 | 0.05 |

TABLE 2
WIYN OBSERVATIONS OF MWC 349A LINE POLARIZATION

| Wavelength (Å) | Line ID | %$P^a$ | $\sigma_P^a$ (%) | PA$^a$ (°) | $\sigma_{PA}^a$ (°) | EW$^a$ (Å) |
|---|---|---|---|---|---|---|
| 5754 | [NII] | 6.96 | 0.32 | 166.5 | 1.3 | 21.4 |
| 5876 | HeI | 7.26 | 0.18 | 161.7 | 0.7 | 38.5 |
| 6562 | Hα | 6.84 | 0.03 | 164.5 | 0.1 | 899.6 |
| 7065 | HeI | 6.18 | 0.12 | 165.0 | 0.5 | 21.9 |
| 7319/7324 | [OII]/[CaII] | 6.21 | 0.01 | 162.6 | 0.4 | 20.9 |
| 8447 | OI | 5.66 | 0.07 | 166.8 | 0.3 | 32.4 |
| 8498/8502 | CaII/Pa16 | 5.10 | 0.15 | 162.4 | 0.9 | 12.1 |
| 8542/8545 | CaII/Pa15 | 4.79 | 0.12 | 158.6 | 0.7 | 14.9 |
| 8599 | Pa14 | 5.13 | 0.21 | 165.0 | 1.2 | 7.0 |
| 8662/8665 | CaII/Pa13 | 4.77 | 0.09 | 158.9 | 0.6 | 22.9 |
| 8750 | Pa12 | 4.93 | 0.16 | 163.4 | 0.9 | 10.5 |
| 8863 | Pa11 | 4.97 | 0.12 | 163.9 | 0.7 | 14.9 |
| 8927 | FeII | 5.16 | 0.23 | 169.7 | 1.3 | 7.0 |
| 9014 | Pa10 | 4.54 | 0.12 | 160.9 | 0.7 | 17.2 |
| 9069 | [SIII] | 4.80 | 0.09 | 163.1 | 0.5 | 27.1 |
| 9229 | Pa9 | 5.22 | 0.06 | 166.3 | 0.4 | 47.4 |
| 9535/9545 | [SIII]/Pa8 | 4.46 | 0.04 | 163.7 | 0.3 | 123.0 |
| 9998 | FeII | 4.84 | 0.15 | 167.0 | 0.9 | 25.3 |
| 10049 | Pa7 | 4.29 | 0.08 | 164.0 | 0.5 | 74.9 |

[a] %$P$, error, PA, and PA error values were calculated from error-weighted means of the Stokes parameters $Q$ and $U$ for the three WIYN observations. Equivalent widths are simple averages over the three observations.

TABLE 3
TIME VARIABILITY OF MWC 349A EMISSION LINE EQUIVALENT WIDTHS

|  |  | EW (Å) | | |
| --- | --- | --- | --- | --- |
| Date | Continuum Flux, Error[a] | OI λ8447 | [SIII] λ9069 | Pa7 λ10049 |
| 2000 Sep 14 | 6 ± 1 | 35.4 | 26.1 | 72.3 |
| 2000 Sep 24 | 36 ± 6 | 32.0 | 27.4 | 74.4 |
| 2000 Sep 25 | 43 ± 7 | 30.8 | 27.7 | 78.1 |

---

[a] Continuum flux was measured at 9754 Å (see §3.2) and is in units of $10^{-14}$ erg cm$^{-2}$ s$^{-1}$ Å$^{-1}$; error was estimated at ~16%, a typical value for a good HPOL observation.

TABLE 4
INTRINSIC POLARIZATION OF MWC 349A

| Date | MJD | %$P$ | $\sigma_P$ (%) | PA (°) | $\sigma_{PA}$ (°) |
|---|---|---|---|---|---|
| | | $V$ | | | |
| 2000 Sep 14 | 1801 | 3.421 | 0.055 | 11.08 | 0.46 |
| 2000 Sep 24 | 1811 | 2.015 | 0.045 | 15.97 | 0.63 |
| 2000 Sep 25 | 1812 | 2.481 | 0.030 | 9.50 | 0.35 |
| | | $R$ | | | |
| 2000 Sep 14 | 1801 | 2.207 | 0.087 | 10.28 | 1.12 |
| 2000 Sep 24 | 1811 | 1.816 | 0.076 | 13.25 | 1.19 |
| 2000 Sep 25 | 1812 | 1.652 | 0.006 | 7.67 | 0.10 |
| | | $I$ | | | |
| 2000 Sep 14 | 1801 | 2.552 | 0.009 | 10.17 | 0.10 |
| 2000 Sep 24 | 1811 | 2.067 | 0.004 | 9.21 | 0.06 |
| 2000 Sep 25 | 1812 | 1.990 | 0.004 | 10.00 | 0.06 |